\documentclass[twocolumn,aps,prl,groupedaddress]{revtex4-2}
\usepackage{graphicx}
\usepackage{amsmath}
\usepackage{color}

\begin{document}

\title{Discrete self-similarity in the formation of satellites for viscous
cavity break-up}
\author{Marco A. Fontelos}
\author{Qiming Wang}
\date{\today}

\begin{abstract}
The breakup of a jet of a viscous fluid with viscosity $\mu _{1}$ immersed
into another viscous fluid with viscosity $\mu _{2}$ is considered in the
limit when the viscosity ratio $\lambda =\mu _{1}/\mu _{2}$ is close to
zero. We show that, in this limit, a transition from ordinary continuous
self-similarity to discrete self-similarity takes place as $\lambda $
decreases. It follows that instead of a single point breakup, the rupture of
the inner jet occurs through the appearance of an infinite sequence of
filaments of decreasing size that will eventually produce infinite sequences
of bubbles of the inner fluid inside the outer fluid. The transition can be
understood as the result of a Hopf bifurcation in the system of equations
modelling the physical problem.
\end{abstract}

\maketitle

\affiliation{Instituto  de  Ciencias  Matem\'aticas,(ICMAT,
  CSIC-UAM-UCM-UC3M),Campus de Cantoblanco,  28006 Madrid, Spain.}


\affiliation{Department of Mathematical Sciences and Center for Applied
Mathematics and Statistics, New Jersey Institute of Technology, Newark, NJ
07102, USA.}










\section{Introduction}

The breakup of fluid filaments into drops is one of the most relevant and
widely studied problems in fluid mechanics. It is both a fundamental problem
since the times of Savart, Plateau and Rayleigh \cite{Savart}, \cite{Plateau}%
, \cite{Rayleigh} and an issue of paramount importance for technological
aplications based on inkjet printing technology, micro and
nano-encapsulation, production of filaments by electrospinning, etc (see
\cite{Eggers} for a general review). The complementary problem, i.e. the
closure of an axisymmetric cavity and the formation of bubbles, has also
been widely studied due to its important role as a mechanism for air
entrainment into the ocean \cite{EF} and also connected to technical
applications \cite{Gordillo}. When a viscous thread or a cavity breaks-up
into drops, it is frequently the case that the result is a sequence of drops
or bubbles of similar size surrounded by droplets/bubbles of much smaller
size called satellite drops/bubbles \cite{GF}, \cite{Th}. The origin and
possible control of such satellites remains a puzzle and an obstacle for
many practical applications where precise knowledge of the size of the
drops/bubbles produced is needed. Remarkably, the pinch-off of a liquid
thread is a singularity \cite{EF2}\ governed by simple similarity laws and
occurs according to selfsimilar functions as first described in \cite{E}. In
this letter we describe and explain the formation of generations of
satellites in the breakup of a very low viscosity fluid jet inside a viscous
environment as the result of a discretely self-similar (DSS) process. This
is in contrast with continuous self-similarity that explains the
disconnection of a single drop from a jet: DSS is a periodic process in
similarity variables, i.e. a process that repeats itself in a logarithmic
time scale (as one approaches a singular time) and decreasing spatial
windows, with the result of infinite sequences of structures of decreasing
size \cite{D}.

\section{Boundary Integral equations and singularity formation}

We consider a viscous jet, with viscosity $\mu _{1}$, inside a viscous
surrounding media with viscosity $\mu _{2}$, and define the viscosity
contrast parameter $\lambda =\frac{\mu _{1}}{\mu _{2}}$. The fluid variables
velocity $\mathbf{v}$ and pressure $p$ for both fluids satisfy Stokes
equations with the condition that the difference of viscous stresses at the
interface balance surface tension (proportional to the interface's mean
curvature $\kappa $, with surface tension coefficient $\gamma $ and along
the normal direction $\mathbf{n}$):%
\begin{equation*}
\left( \mathbf{\sigma }_{2}-\mathbf{\sigma }_{1}\right) \cdot \mathbf{n}%
=\gamma \kappa \mathbf{n}
\end{equation*}
where $\mathbf{\sigma }_{i}$ is the viscous stress exerted by the fluid $i$
on the interface. The fluid interface, described by the distance $h(z,t)$ to
the axis of symmetry, evolves following the velocity field. The interfacial
velocities are solved via boundary-integral equations in axisymmetric
geometry. The for-aft symmetry is imposed so that only half of the drop
shape needs to be determined. The interface velocities are assumed to vary
linearly along each boundary element and integral is computed by using
standard 4- or 6-point Gauss-Legendre formula. The weakly singular integral
is solved by using Gauss-log quadrature. After obtaining velocities, the
interface is updated by the first order Euler time scheme (details can also
be found in \cite{SL}, \cite{W}). When the drop is close to pinching, the
local grid points are redistributed to maintain a smoothly varying spacing
proportional to $\arctan \left( \sqrt{(z-z_{min})^{2}+h_{min}^{2}}\right) $
\cite{BT}. The simulation is stopped when the drop radius is about $10^{-4}$
or smaller (when drop interface oscillation occurs, the simulation may stop
earlier). Validations of the code are presented as supplementary material.

\begin{figure}[tbp]
\centering\includegraphics[width=1.\hsize]{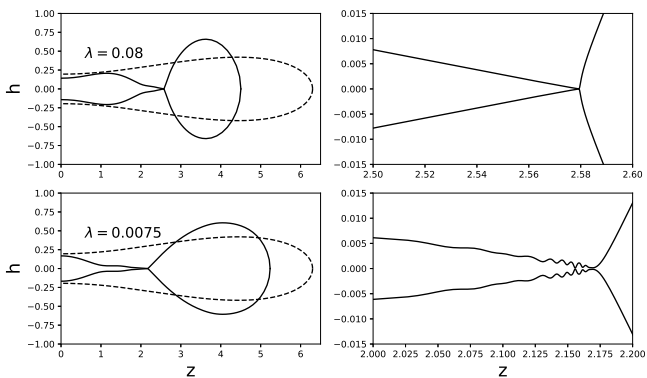}
\caption{ Evolution of a liquid jet and disconnection of a drop. Dotted
line: initial data. Continuous line: finite time breakup through the
formation of \ conical singularities for $\protect\lambda =0.08$ and the
formation of a sequence of satellites for $\protect\lambda =0.0075$. The
right panels are the zoom-in versions of the left panels, at breakup time $%
\tilde{t}\approx 6.065$ and $\tilde{t}\approx 2.025$, respectively. Here $%
\tilde{t}$ is the dimensionless time variable in boundary integral
simulation, before rescaling to long wave variable.}
\label{fig0}
\end{figure}

\begin{figure}[tbp]
\centering\includegraphics[width=1.\hsize]{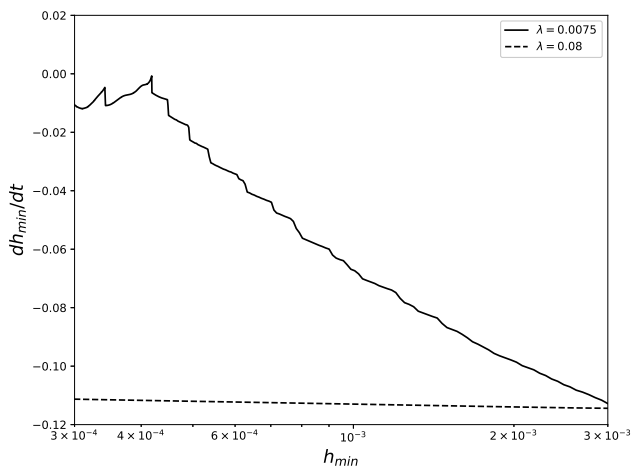}
\caption{ Evolution of $\frac{dh_{\min }}{dt}$, the velocity of the thinning
filament radius vs the minimum radius. According to simple scaling $\frac{%
dh_{\min }}{dt}$ should converge to a constant ad $h_{\min }$ tends to zero.
This is clear for $\protect\lambda =0.08$, but unclear for $\protect\lambda %
=0.0075$. In the later case, oscillations near pinch-off are observed.}
\label{figs}
\end{figure}

\begin{figure}[tbp]
\centering\includegraphics[width=1.\hsize]{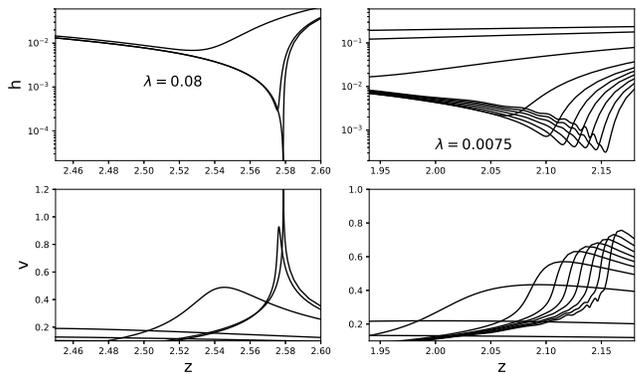}
\caption{ Profiles and axial velocities of the liquid jet for various times
close to the breakup time and showing oscillations in the case $\protect%
\lambda =0.0075$. The profiles for $\protect\lambda=0.08$ correspond to $%
\tilde{t}=6.005, 6.062, 6.065$, while for $\protect\lambda=0.0075$, $\tilde{t%
}= 1.973, 1.992, 2.00, 2.007, 2.013, 2.018, 2.023$.}
\label{figrv}
\end{figure}

Direct scaling yields a linear flow for the filament radius vs time near
pinch-off and a also a linear axial scale vs time: $h(z,t)=(t_{0}-t)\varphi
((z-z_{0})/(t_{0}-t))$ where $t_{0}$,$z_{0}$ are the pinchoff time and point
respectively and $\varphi (\xi )$ is a function behaving linearly at
infinity so that the geometry at pinchoff is that of two cones touching at
their vertex (see figure \ref{fig0}). This is the classical result reported
in the literature \cite{SL2}, \cite{LS}, \cite{W2}. Nevertheless, as we can
appreciate in figures \ref{fig0}, \ref{figs}, for sufficiently small $%
\lambda $, corresponding to an inner fluid of very low viscosity, the
geometry near pinch-off is not as simple as two cones. The evolution of $%
h_{min}$, the minimum jet radius, clearly deviates to the simple linear law
and presents a complex oscillatory behaviour instead. This oscillatory
behaviour translates into both the interface and velocity profiles, as seen
in figure \ref{figrv}, and results in the formation of sequences of
satellites. This is a fact reported in recent experiments \cite{S} in
connection with the healing of an annular viscous film.

\section{Asymptotic model and Discrete self-similarity}

In what follows we study an asymptotic model for the limit of very low
viscosity inner fluid inside an outer viscous fluid. More precisely, by
considering the viscosity ratio $\lambda =\varepsilon ^{2}\lambda _{0}$
where $\lambda _{0}=O(1)$ and $\varepsilon $ is the slenderness ratio
between typical transversal and longitudinal length scales. One can then
obtain the system (see \cite{DC}, \cite{SL2}, \cite{W}):%
\begin{eqnarray}
\frac{\partial h}{\partial t} &=&\frac{1}{16\lambda _{0}}\frac{1}{h}\frac{%
\partial }{\partial z}\left( h^{4}\frac{\partial p}{\partial z}\right)
\label{edo1} \\
p &=&\frac{1}{h}+\frac{2}{h}\frac{\partial h}{\partial t}  \label{edo2}
\end{eqnarray}%
Note that the change of variables $t\rightarrow 2t$, $z\rightarrow z/\sqrt{%
8\lambda _{0}}$ allows to remove the constants $\frac{1}{16\lambda _{0}}$
and $2$ from the equations and, after writing from (\ref{edo2})%
\begin{equation}
\frac{\partial h}{\partial t}=(ph-1)  \label{edo3}
\end{equation}%
and inserting into (\ref{edo1}) we obtain the equation%
\begin{equation}
-\frac{1}{h^{2}}\frac{\partial }{\partial z}\left( h^{4}\frac{\partial p}{%
\partial z}\right) +p=\frac{1}{h}  \label{edo4}
\end{equation}

The system (\ref{edo3}), (\ref{edo4}) consists then of an ordinary
differential equation for $h$ coupled with a linear elliptic partial
differential equation for $p$. We can, in principle, solve the linear
problem (\ref{edo4}) for the pressure $p$ and write%
\begin{equation*}
p=K_{h}\left[ \frac{1}{h}\right]
\end{equation*}%
for some nonlinear integral operator $K_{h}$, and insert into (\ref{edo3})
to obtain%
\begin{equation}
\frac{\partial h}{\partial t}=(K_{h}\left[ \frac{1}{h}\right] h-1)
\label{edo5}
\end{equation}%
which in an integro-differential equation for $h(z,t)$. We will see that,
despite its simplicity (a single first order integro-differential PDE), the
solutions to equation (\ref{edo5}) display a very interesting dynamics
directly related to discretely selfsimilar features (see \cite{D}) leading,
in this case, to the formation of cascades of satellite bubbles of the inner
fluid inside the outer fluid.

\begin{figure}[tbp]
\centering\includegraphics[width=1.\hsize]{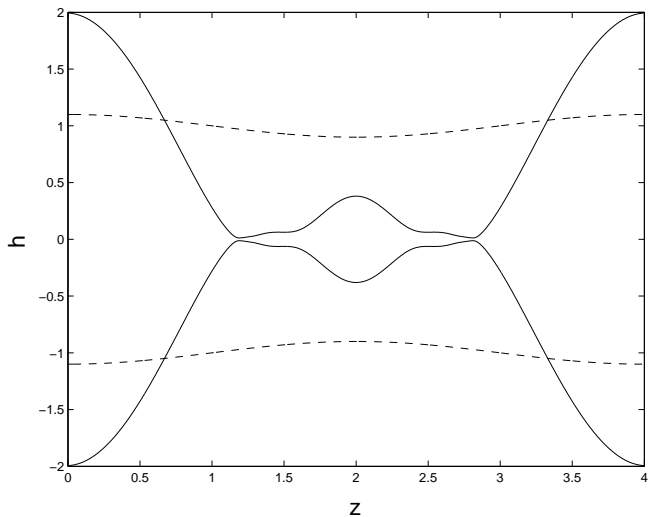}
\caption{ Evolution of a liquid jet. Dotted line: initial data. Continuous
line: finite time breakup through the formation of \ conical singularities.}
\label{fig1}
\end{figure}

We have solved (\ref{edo3}), (\ref{edo4}) numerically for an initial data $%
h(z,0)=1+0.1\cos (2\pi z/L)$ with $L=4$ and obtained the profile in figure %
\ref{fig1} where finite time breakup develops in the apparent form of two
conical shapes touching at the breakup point. This fact can also be observed
in figure \ref{fig3}\ where we represent the evolution profiles for several
times close to the breakup time. The appearance of a seemingly conical
break-up is reinforced by the existence of a selfsimilar solution to (\ref%
{edo3}), (\ref{edo4}) in the form%
\begin{eqnarray*}
h(z,t) &=&(t_{0}-t)H(z/(t_{0}-t)) \\
p(z,t) &=&(t_{0}-t)^{-1}P(z/(t_{0}-t))
\end{eqnarray*}%
where $t_{0}$ is the breakup time and $H(\xi )$, $P(\xi )$ satisfy the system%
\begin{eqnarray}
-H+\xi H^{\prime } &=&HP-1  \label{e1} \\
-\frac{1}{H^{2}}\left( H^{4}P^{\prime }\right) ^{\prime }+P &=&\frac{1}{H}
\label{e2}
\end{eqnarray}

\begin{figure}[tbp]
\centering\includegraphics[width=1.2\hsize]{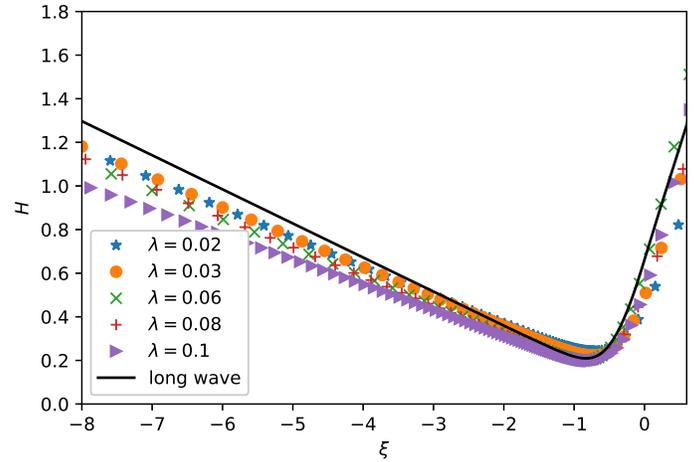}
\caption{Comparison between selfsimilar neck profiles for various values of $%
\protect\lambda$, suitably rescaled, and the selfsimilar solution to the
long wave model. Denoting $\tilde{z}$ and $\tilde{h}$ the dimensionless
variables in full problem, they are connected to the long wave variables as
follows: $\protect\xi = 2(\tilde{z}-\tilde{z}_0)((\tilde{t}_0-\tilde{t})%
\protect\sqrt{8\protect\lambda})^{-1}-b\ln(\tilde{t}_0-\tilde{t})$ and $H =2
\tilde{h}(\tilde{t}_0-\tilde{t})^{-1}$, where $\tilde{z}_0$ and $\tilde{t}_0$
are pinch-off location and time, respectively, in boundary integral
simulation. The constant $b$ is fitted from simulation.}
\label{figprof}
\end{figure}

\begin{figure}[tbp]
\centering\includegraphics[width=1.\hsize]{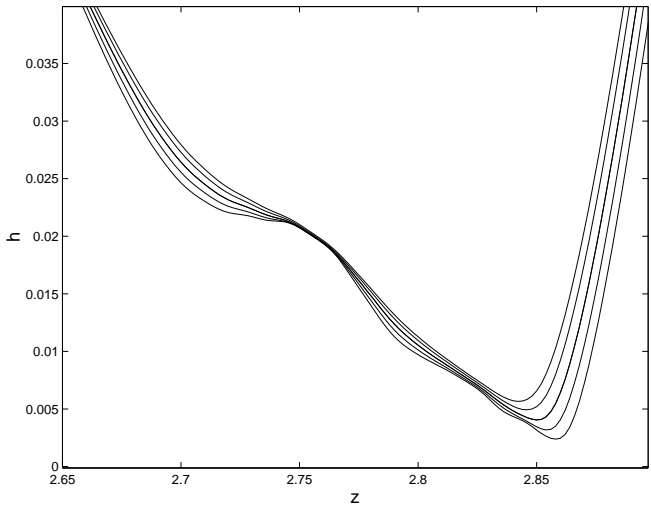}
\caption{Numerical profiles for five different times as we approach the
singularity. The singularity time is $t_0=3.970$ and the profiles correspond
to $t=3.940,3.945,3.950,3.955,3.960$.}
\label{fig2}
\end{figure}

\begin{figure}[tbp]
\centering\includegraphics[width=1.\hsize]{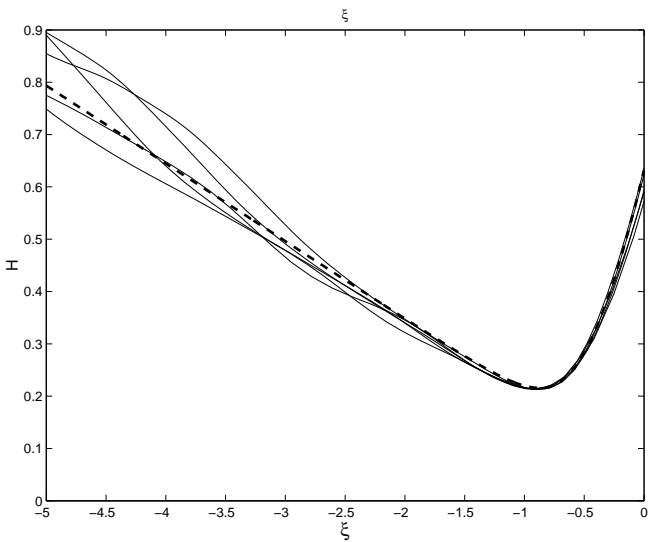}
\caption{The profiles of \protect\ref{fig2} rescaled according to the
similarity laws together with the selfsimilar profile.}
\label{fig3}
\end{figure}

together with the boundary conditions $H\sim \nu _{\pm }\xi $, $P\sim \mu
_{\pm }\xi ^{-1}$ as $\xi \rightarrow \pm \infty $. Note that the linear
asymptotic behaviour of $H$ implies that the local shape near the breakup
point is in the form of two cones with semi angles $\theta _{\pm }=\arctan
\nu _{\pm }$. By means of a Newton iteration scheme we have solved (\ref{e1}%
), (\ref{e2}) subject to boundary conditions implying a linear asymptotic
behaviour of $H(\xi )$ and found the profile represented in dotted lines in
figure \ref{fig2}. In particular, we can compute $\theta _{+}\simeq 44%
{{}^o}%
$ and $\theta _{-}\simeq 9%
{{}^o}%
$. It is worth noting the remarkable good agreement between the selfsimilar
solution to the long wave model and the selfsimilar profiles obtained from
the full boundary integral calculation (figure \ref{figprof}). In order to
compare them we rescale time by a factor of $2$ and $z$ by a factor $\sqrt{%
8\lambda }=\sqrt{8\lambda _{0}}\varepsilon $ also see caption in Fig. \ref%
{figprof} , in agreement with the rescaling that led to (\ref{edo3}), (\ref%
{edo4}). We can observe in figure \ref{figprof} that, very interestingly,
the resulting profiles seem to converge towards a universal profile as $%
\lambda $ decreases and that profile is very close to the selfimilar
solution to the long wave model. Nevertheless, as we mentioned above, the
comparison cannot hold for artibrarily small values of $\lambda $ due to the
development of instabities around the pinch-off point. Below we will present
a detailed study of this instabilities in the context of the long wave model.

\begin{figure}[tbp]
\centering\includegraphics[width=1.\hsize]{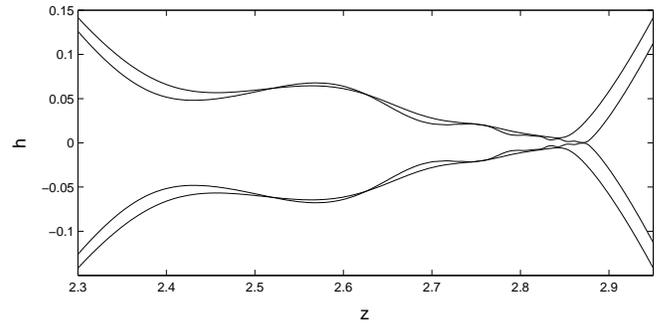}
\caption{Two profiles close to the singularity time. The one closer to $%
t_{0} $ has developed a sequence of undulations in the form of iterated
structures approaching the point of breakup.}
\label{fig4}
\end{figure}

\begin{figure}[tbp]
\centering\includegraphics[width=1.\hsize]{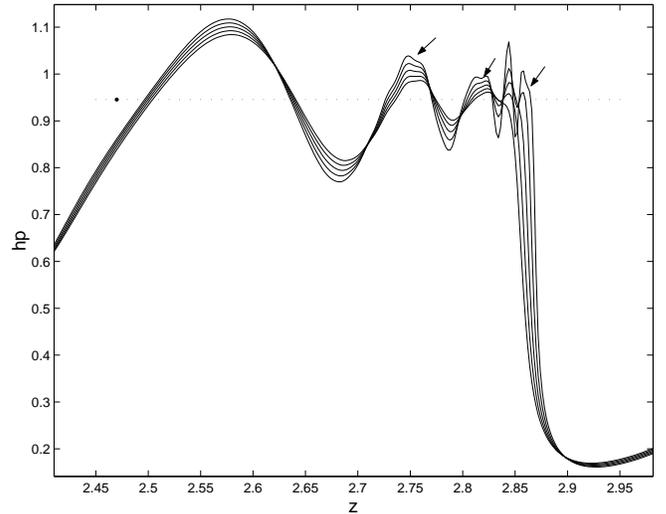}
\caption{The function $hp$ for various times close to $t_{0}$ where the
presence of a sequence of oscillations is visible. Also, indicated with
arrows, the development of secondary oscillations that will give rise to the
appearance of cascades of secondary structures.}
\label{fig5}
\end{figure}

In figure \ref{fig3} we plot the same profiles as in figure \ref{fig2}
rescaled by the factor $(t_{0}-t)^{-1}$ and compared to the selfsimilar
solution. As we can see, the numerical profiles evolve around the
selfsimilar solution but do not seem to converge to it. In fact, as we
approach the singularity, instabilities develop in the form of oscillations
whose length-scale decreases as we approach the singularity point (see the
undulations in figure \ref{fig3}). This is the signature of discrete
self-similarity (DSS), i.e. $\tau $-periodic solutions of the equations (\ref%
{edo3}), (\ref{edo4}) in similarity variables $\tau =-\log (t_{0}-t)$, $\xi
=z/(t_{0}-t)$:%
\begin{eqnarray}
H_{\tau }-H+\xi H^{\prime } &=&HP-1 \\
-\frac{1}{H^{2}}\left( H^{4}P^{\prime }\right) ^{\prime }+P &=&\frac{1}{H}
\end{eqnarray}%
As we showed in \cite{D}, discretely self-similar solutions may appear as
the result of a Hopf bifurcation around a continuously selfsimilar solution
where the later solution becomes unstable giving rise to a periodic orbit
around it (the DSS solution). In the present case, our numerical results
show evidence of periodicity in the self-similar variables and the result is
an infinite sequence of iterated filaments of decreasing length-scale that
develop as we approach the singularity (see figure \ref{fig4}). The breakup
of these filaments can give rise to and infinite sequence of satellite
bubbles of the inner fluid inside the outer fluid. The presence of periodic
structures is also visible in \ref{fig5}, where we represent $hp$ which,
according to the scaling laws for $h$ and $p$ near the singularity, should
converge to a universal function $HP$ that tends to constant $\nu _{\pm }\mu
_{\pm }$ as $\left\vert \xi \right\vert \rightarrow \infty $. As we can see
in the figure, the behaviour is not that of convergence to a constant
(dotted line) but oscillatory around that constant instead. Secondary
instabilities, visible in figure \ref{fig5}, will develop into secondary
cascades of iterated filaments and hence may give rise to new cascades of
subsatellite bubbles.

\begin{figure}[tbp]
\centering\includegraphics[width=90mm,height=50mm]{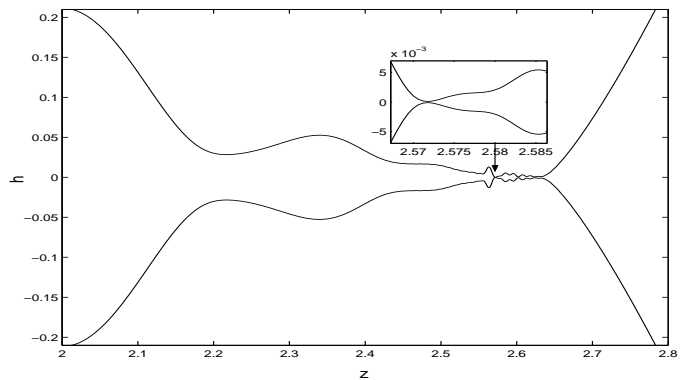}
\caption{Geometry of a jet near breakup for $g=0$. A sequence of bubbles can
be be seen together with the detail near breakup in the neighborhood of a
satellite. }
\label{fig7}
\end{figure}

In order to obtain a better understanding of the origin of DSS solutions in
this problem, we consider now higher order viscous effects for the pressure
in (\ref{edo2}), in the form (cf. \cite{W})
\begin{equation*}
p=\frac{1}{h}+\frac{2}{h}\frac{\partial h}{\partial t}+\frac{\varepsilon
^{2}\left\vert \log \varepsilon \right\vert }{h^{2}}\frac{\partial }{%
\partial z}\left( h^{4}\frac{\partial }{\partial z}\frac{1}{h}\right)
\end{equation*}%
(see the deduction in Supplementarial material). This leads, after the same
rescaling in $z$ and $t$ leading to (\ref{edo3}), (\ref{edo4}), to the
system formed by (\ref{edo4}) and%
\begin{equation*}
\frac{\partial h}{\partial t}=(ph-1)+\frac{g}{h}\frac{\partial }{\partial z}%
(h^{2}h_{z})
\end{equation*}%
where $g\simeq 4\lambda _{0}\varepsilon ^{2}\left\vert \log \varepsilon
\right\vert $. Numerically we found a transition between damped oscillations
and periodic solutions (a Hopf bifurcation, as shown in the stability
analysis provided in the Supplementary material) when the parameter $g\simeq
3\cdot 10^{-4}$. A periodic orbit in self-similar coordinates implies that
the breakup of the jet occurs in such a way that self-similarity holds at
discrete times $t_{1}$, $t_{2}$,..., approaching the breakup time $t_{0}$;
if $T$ is the period of the orbit in selfsimilar coordinates, then $%
t_{N+1}/t_{N}=e^{-T}$. Then, the sequence of bubbles that appear are located
at distances $d_{1}$, $d_{2}$,...,$d_{N}$ from $z_{0}$ (with $%
d_{N}\rightarrow 0$ as $N\rightarrow \infty $) and will have a radius $r_{1}$%
, $r_{2}$,...,$r_{N}$,.... Using $h=e^{-\tau }H$ and $z-z_{0}=e^{-\tau }\xi $
we deduce
\begin{equation*}
\frac{r_{N+1}}{r_{N}}=e^{-T}\text{, }\frac{d_{N+1}}{d_{N}}=e^{-T}\ .
\end{equation*}%
Away from the Hopf bifurcations (at $g=0$, for instance) secondary
instabilities may create new cascades of bubbles between consecutive bubbles
which may again contain new cascades, etc, creating a fractral set. In
figure \ref{fig7} we represent the result, near breakup, of the evolution
for initial data $h(z,0)=1+\frac{3}{4}\cos \left( \frac{\pi z}{2}\right) $
and for $g=0$. A sequence of bubbles can be be appreciated and breakup will
occur near a satellite bubble. The region around the breakup point presents
the characteristic two-cone geometry plus instabilities, as in figure \ref%
{fig4}, which will give rise to new cascades of bubbles, and the process
will repeat infinite times.

To summarize, we have shown the mechanism for the appearance of satellites
for a viscous jet immersed into another viscous fluid with a much higher
viscosity. The study of similar mechanisms for the appearance of cascade of
structures (filaments) produced in very viscous jets \cite{SB} and the
iterated stretching of viscoelastic filaments \cite{OM} is underway.

\begin{acknowledgments}
  M.A. Fontelos is supported by the research Grant from the Spanish Ministry of Economy and Competitiveness MTM2017-89423-P.

\end{acknowledgments}

\end{document}